\documentclass[prl,twocolumn]{revtex4}
\usepackage{amsfonts}
\usepackage{amsmath}
\usepackage{amssymb}
\usepackage{graphicx}
\usepackage[usenames]{color}%

\begin{document}

\title{Consistency of field-theoretical and kinetic calculations of 
viscous transport coefficients for a relativistic fluid}
\author{Gabriel S.\ Denicol$^1$}
\author{Xu-Guang Huang$^{1,2}$}
\author{Tomoi Koide$^2$}
\author{Dirk H.\ Rischke$^{1,2}$}
\affiliation{$^1$ Institut f\"ur Theoretische Physik, 
Johann Wolfgang Goethe-Universit\"at, D-60438 Frankfurt am Main, Germany\\
$^2$ Frankfurt Institute for Advanced Studies, 
D-60438 Frankfurt am Main, Germany}
\begin{abstract}
The transport coefficients of causal relativistic 
dissipative fluid dynamics are 
calculated both in a field-theoretical and a kinetic approach.
We find that the results from the traditional kinetic calculation 
by Israel and Stewart are modified. The new expressions for the
viscous transport coefficients agree with the results obtained 
in the field-theoretical approach when the contributions from pair 
creation and annihilation are neglected.
\end{abstract}

\maketitle

Relativistic fluid dynamics is an important model to understand
various collective phenomena in astrophysics and heavy-ion collisions. 
However, the relativistically covariant extension of the 
Navier-Stokes equations is acausal
and unstable \cite{dkkm3}. The reason is that
the irreversible currents (the shear stress tensor $\pi^{\mu\nu}$, 
the bulk viscous pressure $\Pi$ etc.) are linearly proportional
to the thermodynamic forces (the shear tensor $\sigma^{\mu \nu}$,
the expansion scalar $\theta$ etc.), with the constant
of proportionality being the shear viscosity coefficient $\eta$,
the bulk viscosity coefficient $\zeta$ etc.. Thus, the forces have
an instantaneous influence on the currents, which obviously violates
causality and leads to instabilities.
These problems are solved by introducing retardation into the 
definitions of the irreversible currents, leading to
equations of motion for these currents which thus become
independent dynamical variables. Theories of this type are called
causal relativistic dissipative fluid dynamics (CRDF).

With the retardation, in general the irreversible currents and the 
thermodynamic forces are no longer linearly proportional 
to each other.
As a consequence, the transport coefficients for CRDF cannot be computed
applying the methods commonly used for Navier-Stokes fluids, 
such as the Green-Kubo-Nakano (GKN) formula.
So far, there are several approaches to derive the 
transport coefficients of CRDF 
\cite{is,muronga,koide1,koide2,knk,hkkr,moore,conformal,b}.
They were first calculated by Israel and Stewart (IS) 
applying the so-called
14-moment approximation to the Boltzmann equation \cite{is,muronga}.
However, the Boltzmann equation is applicable
only in the dilute limit and hence we cannot expect
these results to describe the behavior of dense fluids.
Recently, a new microscopic formula 
to calculate the transport coefficients of CRDF  
from time-correlation functions was proposed \cite{koide1,koide2,knk,hkkr}. 
This formula is the analogue of the GKN formula in Navier-Stokes fluids.
Since this formula is derived from quantum field theory, 
it will be applicable even to dense fluids.

However, in a leading-order perturbative calculation
which should apply in the dilute limit, i.e., the regime of applicability
of the kinetic approach, the field-theoretical formula gives 
results which
are different from those of the IS calculation.
Is this inconsistency due to a problem with 
the field-theoretical formula or with the kinetic calculation?
Or is there simply no correspondence between
the field-theoretical and the kinetic derivation of
the transport coefficients of CRDF? This would come as a
surprise, as this correspondence does exist in the case of 
Navier-Stokes fluids.

In this letter, we show that the field-theoretical 
and kinetic calculations are indeed consistent, even for CRDF.
The key point is that the 14-moment approximation 
employed by Israel and Stewart is not unique.
We suggest a new method to obtain equations of motion
for the irreversible currents, which leads to
expressions for the transport coefficients which are different from
the IS results. The new transport coefficients
turn out to be consistent with those obtained from the field-theoretical 
formula.

In the original IS calculation, the evolution equations of the shear stress 
tensor and the bulk viscous pressure are obtained from the 
second moment of the Boltzmann equation,
\begin{equation}
\partial_{\mu} \int\frac{d^3{\bf k}}{(2\pi)^3 E_{\bf k}} 
\,K^{\mu}K^{\nu}K^{\rho}\, f 
= \int\frac{d^3{\bf k}}{(2\pi)^3 E_{\bf k}}\, K^{\nu}K^{\rho}\,C[f],
\label{2ndmomentBE}
\end{equation}
where $f$ is the single-particle distribution function,
$K^{\mu} = (E_{\bf k},{\bf k})$ with 
$E_{\bf k} = \sqrt{{\bf k}^2 + m^2}$, and $C[f]$ 
is the collision term.
In order to obtain a closed set of equations, 
one assumes a specific form for $f$,
\begin{equation} \label{f}
f = f_0 + f_0 (1-af_0)(e + e_{\mu}K^{\mu} + e_{\mu\nu}K^{\mu}K^{\nu}),
\end{equation}
where $e$, $e^{\mu}$, and $e^{\mu\nu}$ constitute a set
of 14 independent parameters related to the irreversible currents
by matching conditions and $f_0 = ( e^{\beta u_{\mu} K^{\mu}} + a)^{-1}$ is
the single-particle distribution function in local
equilibrium, with $a=\pm 1$ for fermions/bosons;
$\beta \equiv 1/T$ is the inverse temperature. 
Then Eq.\ (\ref{2ndmomentBE}) is decomposed into scalar, vector, 
and tensor parts which are interpreted as 
the evolution equations of the bulk viscous pressure, 
the particle diffusion (heat conduction) current, and 
the shear stress tensor, respectively.

The idea of the new kinetic calculation is as follows \cite{dhknr}.
In the kinetic approach, the shear stress tensor and the 
bulk viscous pressure are always defined by 
\begin{eqnarray}
\pi^{\mu\nu} 
&=& \Delta^{\mu\nu}_{\hspace*{0.3cm}\alpha\beta} 
\int \frac{d^3{\bf k}}{(2\pi)^3 E_{\bf k}} K^{\alpha}K^{\beta} (f - f_0), \\
\Pi
&=& -\frac{m^2}{3} \int \frac{d^3{\bf k}}{(2\pi)^3 E_{\bf k}} (f-f_0)\;.
\end{eqnarray}
The tensor $\Delta^{\mu\nu\alpha\beta} 
= (\Delta^{\mu\alpha}\Delta^{\nu\beta} + 
\Delta^{\mu\beta}\Delta^{\nu\alpha} 
- (2/3) \Delta^{\mu\nu} \Delta^{\alpha\beta})/2$,
with $\Delta^{\mu\nu} = g^{\mu\nu} -u^{\mu}u^{\nu}$.
The evolution equations for $\pi^{\mu\nu}$ and $\Pi$ are now
obtained directly by applying the comoving time derivative 
and substituting the Boltzmann equation together with Eq.\ (\ref{f}).
Details will be presented elsewhere \cite{dhknr}.

The final result for the evolution equations is
\begin{eqnarray}
&&\hspace{-1cm} \Delta^{\mu\nu}_{\hspace*{0.3cm}\alpha\beta} 
u^{\rho}\partial_{\rho} \pi^{\alpha \beta} 
= 
- \frac{\pi^{\mu\nu}}{\tau_{\pi}}
+ 2 (\beta_{\eta} + \eta_{\pi\Pi} \Pi) \sigma^{\mu\nu} \nonumber \\
&& \hspace{-1cm}- 2 \eta_{\pi\pi} \Delta^{\mu\nu}_{\alpha\beta} 
\pi^{\alpha}_{\lambda}\sigma^{\beta\lambda} 
+ 2 \Delta^{\mu\nu}_{\alpha\beta} \pi^{\alpha}_{\lambda}
\omega^{\beta\lambda} 
+ \eta_{\pi\theta} \pi^{\mu\nu} \theta, \label{shear}\\ 
&&\hspace{-1cm} u^{\rho}\partial_{\rho} \Pi 
= - \frac{\Pi}{\tau_{\Pi}}
-(\beta_{\zeta} + \zeta_{\Pi\Pi} \Pi) \theta
+ \zeta_{\Pi \pi} \pi^{\mu\nu}\sigma_{\mu\nu} , \label{bulk}
\end{eqnarray}
where $\nabla^{\mu} = \Delta^{\mu\nu}\partial_{\nu}$, 
$\theta = \partial_\mu u^\mu$ is the expansion scalar,
$\sigma^{\mu\nu} = \Delta^{\mu\nu}_{\hspace*{0.3cm}\alpha\beta}
\nabla^{\alpha} u^{\beta}$ is the shear tensor,
and $\omega^{\mu\nu} = (\nabla^{\mu}u^{\nu} - \nabla^{\nu}u^{\mu})/2$
is the vorticity.
The relaxation times for shear and
bulk viscous pressure are $\tau_\pi$ and $\tau_\Pi$, respectively,
and the transport coefficients $\beta_\eta \equiv \eta/\tau_\pi$ and
$\beta_\zeta \equiv \zeta/\tau_\Pi$.
The other coefficients in Eqs.\ (\ref{shear}),
(\ref{bulk}) play no role in the following discussion.

In general, the values of the transport coefficients 
depend on the collision term.
However, in the ratios $\beta_\eta,\, \beta_\zeta$
of viscosity coefficients to relaxation times the
collision term drops out; these ratios
are simply thermodynamic functions
\begin{eqnarray}
\beta_{\eta} &=& \frac{1}{15} \left[ 9\,P + \varepsilon 
- m^4 \int \frac{d^3 {\bf k}}{(2 \pi )^3 E^3_{\bf k}}\,
f_0({\bf k}) \right], \label{be} \\
\beta_{\zeta} &=& \left( \frac{1}{3} -  c^2_s \right) 
(\varepsilon + P) 
- \frac{2}{9}\,(\varepsilon - 3 P)  \nonumber \\
&&- \frac{m^4}{9} \int \frac{d^3 {\bf k}}{(2\pi)^3 E^3_{\bf k}}\,
f_0 ({\bf k})\;. \label{bz}
\end{eqnarray}
Here, $\varepsilon$, $P$, and $c^2_s = dP/d \varepsilon$ 
are the energy density, the pressure, 
and the velocity of sound squared, respectively.
In the following, we shall prove that the values (\ref{be}) and (\ref{bz})
obtained in the new kinetic calculation are consistent
with the field-theoretical approach.

The field-theoretical approach uses the
projection operator method \cite{koide1,koide2,knk,hkkr}.
First, we discuss the shear viscosity.
The expression for $\beta_\eta$ is
\begin{equation}
\beta_{\eta, {\rm ft}} = (\varepsilon + P) \,
\frac{\int d^3{\bf x}\,(T^{yx}({\bf x}),T^{yx}({\bf 0}))}{
\int d^3{\bf x}\,(T^{0x}({\bf x}),T^{0x}({\bf 0}))} \;, \label{betae}
\end{equation} 
where $T^{\mu\nu}$ is the energy-momentum tensor and the inner product 
is defined by Kubo's canonical correlation, 
$(X,Y)=\int_{0}^{\beta}d\lambda\, \beta^{-1}
\langle e^{\lambda H}Xe^{-\lambda H}Y \rangle_{\rm eq}$. 
Here, $H$ is the Hamiltonian and $\langle \cdots \rangle_{\rm eq}$
indicates the thermal expectation value.

In order to compare with kinetic theory, 
it is sufficient to calculate this correlation functions appearing
in Eq.\ (\ref{betae}) in the free gas approximation.
Using $\int d^3{\bf x}\,(T^{0x}({\bf x}),T^{0x}({\bf 0})) 
= (\varepsilon + P)/\beta$, we obtain for bosons
\begin{eqnarray}
\beta_{\eta, {\rm ft}} & & = -\beta \frac{\partial}{\partial \beta} 
\int \frac{d^3{\bf k}}{(2\pi)^3}\, \frac{({\bf k}^2)^2}{15E^3_{\bf k}}\,
 f_0 ({\bf k}) \nonumber \\
&& + \int \frac{d^3{\bf k}}{(2\pi)^3}\, \frac{({\bf k}^2)^2}{30E^3_{\bf k}}\,
[1 + 2\, f_0 ({\bf k})]\;. \label{e/t_full}
\end{eqnarray}

The first term in Eq.\ (\ref{e/t_full}) 
contains the derivative with respect to $\beta$, which is re-expressed using
\begin{eqnarray}
\frac{\partial f_0({\bf k})}{\partial \beta}
&=& \frac{E_{\bf k}}{\beta} \nonumber \\
&& \hspace{-2.4cm}\times \lim_{{\bf p} \rightarrow 0} 
\frac{f_0({\bf k+p})[1+f_0({\bf k})] - f_0({\bf k})[1+f_0({\bf k+p})]}
{E_{\bf k+p} - E_{\bf k}}\;.
\end{eqnarray}
In many-body physics \cite{ftemp}, this term can be interpreted as 
the contribution from the collisions of bosons.
The second term in Eq.\ (\ref{e/t_full}) 
contains an ultraviolet divergent term due to the vacuum self-energy 
and is re-expressed as
\begin{eqnarray}
&& 1 + 2\,f_0({\bf k}) \\
&= & \lim_{{\bf p} \rightarrow 0}
\{ [1+f_0({\bf k+p})] [1+f_0({\bf k})] - f_0({\bf k+p}) f_0({\bf k}) \}\;.
\nonumber
\end{eqnarray}
This term corresponds to the pair annihilation-creation (PAC) part.

Usually, pair annihilation and creation processes are
not considered in the Boltzmann equation.
Thus, for the sake of a consistent comparison, 
we compare only the collision part in Eq.\ (\ref{e/t_full}) 
with the kinetic result. A straightforward integration by parts
gives the result (\ref{be}), i.e., it is identical with the
result of the new kinetic calculation.
The full result, i.e., including the PAC part, is 
\begin{equation}
\beta_{\eta, {\rm ft}} = P\, , \label{shear_b}
\end{equation}
which has already been quoted in Ref.\ \cite{knk}. Here,
the vacuum contribution has been neglected.

\begin{figure}[t]
\includegraphics[scale=0.3]{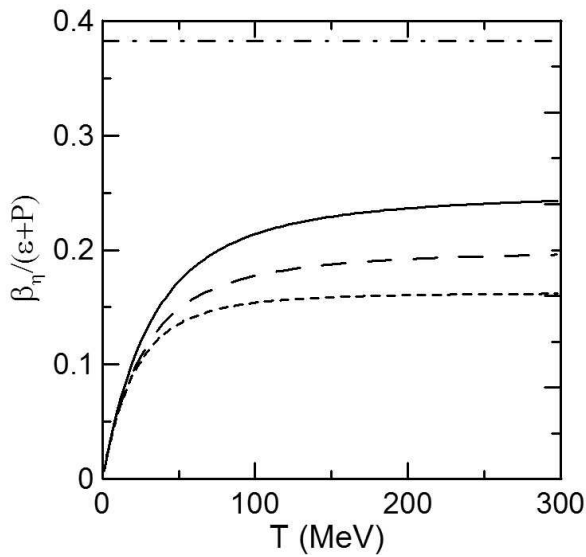}
\caption{The temperature 
dependence of $\beta_\eta/(\varepsilon+P)$ for pions, $m=140$ MeV. 
The solid, dashed, and dotted lines 
represent the results of field-theoretical approach including the PAC part, 
the new kinetic calculation, and the IS calculation, respectively.
For the sake of comparison, the result from an AdS/CFT calculation,
$\beta_\eta/(\varepsilon+P) = 1/[2 (2-\ln 2)]$ \cite{conformal},
is shown by the dash-dotted line.}
\label{fig1}
\end{figure}

The temperature dependence of $\beta_\eta/(\varepsilon +P)$
is shown in Fig.\ \ref{fig1}.
One observes that the values obtained from the field-theoretical formula 
are larger than those for the new kinetic and the IS calculation. 
The reason is that the PAC part makes a non-negligible contribution 
to $\beta_{\eta,{\rm ft}}$.
But even without this part, the new kinetic calculation gives values
which are above the original IS result.
Since $\beta_\eta/(\varepsilon +P)$ is related to the signal propagation 
speed in CRDF, we expect effects on the collective behavior of 
relativistic fluids \cite{dkkm3}.    
The same transport coefficients of CRDF were calculated from the 
Boltzmann equation in Ref.\ \cite{moore}, 
but the results depend on the coupling strength and are different 
from our results.

Next, we consider $\beta_\zeta$.
In the field-theoretical approach, this is given by 
\begin{equation} 
\beta_{\zeta, {\rm ft}} = (\varepsilon +P)\,
\frac{ \int d^3{\bf x}\, (\hat{\Pi} ({\bf x}),\hat{\Pi}({\bf 0}))}{
\int d^3{\bf x}\,(T^{0x}({\bf x}),T^{0x}({\bf 0}))} \;, \label{betaz}
\end{equation}
where $\hat{\Pi} = [(1 - 3\, c^2_s) \,T^{00}
- T^{\mu}_{\hspace*{0.15cm} \mu}]/3$ \cite{hkkr}.
For bosons in the free gas approximation, we obtain
\begin{eqnarray}
\beta_{\zeta,{\rm ft}} & = & -\beta 
\int \frac{d^3 {\bf k}}{(2\pi)^3} \frac{1}{9 E^3_{\bf k}}
\left( {\bf k}^2  - 3\, c^2_s E_{\bf k}^2 \right)^2 
\frac{\partial}{\partial \beta}  f_0({\bf k}) \nonumber \\
&+ & \frac{m^4}{18} \int \frac{d^3{\bf k}}{(2\pi)^3  E^3_{\bf k}}\, 
[1+2\, f_0({\bf k})]\;. \hspace*{3cm}
\end{eqnarray}
Similarly to the case of $\beta_{\eta, {\rm ft}}$, 
the first and second terms 
are interpreted as the collision part and the PAC part, respectively.
The collision part can be re-expressed using
integration by parts, thermodynamic relationships, and the definition
of $c_s^2$. It is then found to be exactly the same as
the new kinetic result (\ref{bz}).

By taking the PAC part into account, the 
field-theoretical approach yields
\begin{equation}
\beta_{\zeta,{\rm ft}} = 
\left( \frac{1}{3} - c^2_s \right) (\varepsilon + P) - 
\frac{2}{9}(\varepsilon - 3 P)\;. \label{bulk_b}
\end{equation}
The detailed derivation will be given in Ref.\cite{hkkr}.

\begin{figure}[t]
\includegraphics[scale=0.3]{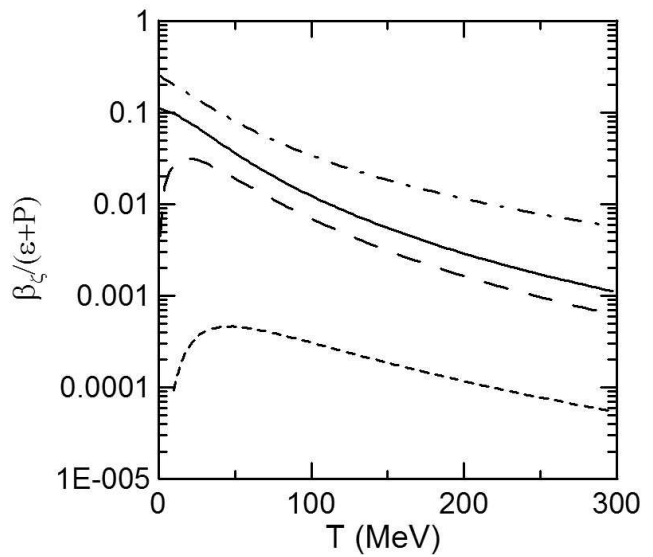}
\caption{The temperature
dependence of $\beta_\zeta/(\varepsilon +P)$ for pions, $m=140$ MeV. 
The solid, dashed, and dotted lines 
represent the results of field-theoretical approach
including the PAC part, the new kinetic calculation,
and the IS calculation, respectively.
For the sake of comparison, the result from a string theoretical
calculation \cite{b} is shown by the dash-dotted line.}
\label{fig2}%
\end{figure}

The temperature dependence of $\beta_\zeta/(\varepsilon + P)$ 
is shown in Fig.\ \ref{fig2}.
Similarly to the case of the shear viscosity, 
the new kinetic calculation and the field-theoretical formula predict 
larger values than the IS calculation.
However, the difference is now more than an order of magnitude.

For $\beta_\zeta$ the PAC part qualitatively 
changes the behavior, in particular, at low temperatures.
If we neglect the PAC part, 
$\beta_\zeta$ increases rapidly 
in the low-temperature region, and starts to decrease at high temperature.
This behavior is the same in all kinetic calculations.
However, when we consider the PAC part, $\beta_\zeta$ becomes 
a monotonically decreasing function of temperature.
This is the same as the behavior predicted by a string theoretical
calculation \cite{b}.

In the case of the kinetic calculations, 
$\beta_\eta$ and $\beta_\zeta$
have the same forms for fermions as for bosons.
This is true even for the field-theoretical approach,
if the PAC part is neglected.
Then, $\beta_\eta$ and $\beta_\zeta$ for fermions 
are given by Eqs.\ (\ref{be}) and (\ref{bz}), respectively,
i.e., the field-theoretical approach and the new kinetic calculation 
yield the same results.
However, the contributions from the PAC terms depend on the statistics.
For fermions,  
the full results of the field-theoretical calculation are
\begin{eqnarray}
\beta_{\eta,{\rm ft}}& = & 0\;, \\
\beta_{\zeta, {\rm ft}} & = & 
\left( \frac{1}{3} - c^2_s \right) (\varepsilon + P) 
- \frac{1}{3}\,(\varepsilon - 3 P)\;.
\end{eqnarray}
The results of the field-theoretical calculation may be
expressed in a unified way as
\begin{eqnarray}
\beta_{\eta, {\rm ft}}
&=& |3-\alpha|\,P\; ,  \label{uni_e}\\
\beta_{\zeta, {\rm ft}}
&=& 
\left( \frac{s}{3}\frac{d}{d s} - \frac{\alpha}{9} \right) 
(\varepsilon - 3 P)\; , \label{uni_z}
\end{eqnarray}
where $\alpha=2$ for boson and $\alpha=3$ for fermion, 
and $s$ is the entropy density.
Note that, for a mixed system of bosons and fermions, 
the correlation functions which appear in the numerators and denominators 
of Eqs.\ (\ref{betae}) and (\ref{betaz}) 
are the sum of both contributions, respectively.

Note that these calculations 
are leading-order results and will be modified by the effect of interactions.
For example, the exact expression for $\tau_\pi /\beta$ 
is given by the ratio of the real and imaginary parts 
of the retarded Green's function of $T^{yx}$ \cite{hkkr}.
However, in the leading-order calculation, 
the real part is approximated by the result for the free-gas approximation,
while the imaginary part is not.
The divergent $\tau_{\pi}$ for fermions, leading to a
vanishing $\beta_{\eta, {\rm ft}}$ 
in Eq.\ (\ref{uni_e}), will be rendered finite
by a more complete calculation.

As shown in Ref.\ \cite{sr}, there is a sum rule for the 
bulk viscous pressure. There, the correlation function for
bulk viscous pressure was calculated for 
interacting gauge bosons. In the weak-coupling limit, the result 
is reproduced by
setting $\alpha = 4$ in Eq.\ (\ref{uni_z}).
Similar correlation functions were studied in Lattice QCD \cite{mey}.

Now we can answer the question posed in the introduction.
The inconsistency between the field-theoretical
and kinetic calculations is due to a problem with the IS calculation.
We have suggested an alternative kinetic calculation, 
the results of which are consistent with the field-theoretical approach.
Thus, even in CRDF, there is a relation between the field-theoretical 
and kinetic calculations, just as for Navier-Stokes fluids.

At the same time, we found that, in the field-theoretical formula, 
there is a contribution from the PAC part which is not included 
in the kinetic calculation. In a relativistic setting, particle
annihilation and creation processes may occur, which re-distribute
momenta as well as influence chemical equilibrium, thus affecting
both the shear and the bulk viscosities. Therefore,
one must not neglect the PAC part.
When taking the PAC part into account, we have seen that
$\beta_{\eta, {\rm ft}}$ and $\beta_{\zeta, {\rm ft}}$ can be expressed 
solely by thermodynamic quantities such as 
$\varepsilon$, $P$, and $c_s$ as shown in Eqs.\ 
(\ref{uni_e}) and (\ref{uni_z}).

A natural question that arises from our work is 
whether it is possible to introduce the PAC term
in the kinetic calculation.
Remember that $\beta_\eta$ and $\beta_\zeta$ are independent of the 
collision term.
Thus, even if we improve the collision term by introducing PAC
processes, these coefficients will not change.
Therefore we have to extend the Boltzmann equation itself to 
include the PAC effect discussed here.

In this letter, we simply dropped the temperature-independent, 
divergent vacuum contribution term in the calculation of the 
coefficients. Strictly speaking, however, we do not know 
the precise renormalization scheme for these quantities.
The problem of renormalization is very important in order 
to reliably determine the values of the transport coefficients and 
should be studied more carefully in the future.

We have discussed the consistency of the 
field-theoretical and kinetic calculations.
We have also computed the 1+1-dimensional scaling
flow using our new kinetic coefficients and compared with 
a numerical simulation of the Boltzmann equation. We 
find better agreement with the new coefficients than
with those of the IS calculation \cite{dhknr}. 
This may serve as another justification for the validity of our results.

In order to discuss particle diffusion (heat conduction), 
the kinetic approach should be generalized to a 
multi-component fluid, considering the flows 
of particles and anti-particles on an equal footing.
For the field-theoretical approach, it is not yet known what is 
the appropriate projection operator to 
derive the corresponding formula for the particle diffusion.
This is a challenge for the future.

T.K.\ would like to thank
H.\ Abuki, D. Fernandez-Fraile, T.\ Hatsuda, Y.\ Hidaka, T. Kodama, J.\ Noronha and A. Peshier 
for fruitful discussions and comments.
This work was
(financially) supported by the Helmholtz International Center
for FAIR within the framework of the LOEWE program (Landesoffensive zur
Entwicklung Wissenschaftlich- \"Okonomischer Exzellenz) 
launched by the State of Hesse.

\end{document}